\documentclass{article}
\usepackage{amsmath,amssymb,theorem,cite}

\newtheorem{theorem}{Theorem}[section]   

\newtheorem{proposition}[theorem]{Proposition}

\numberwithin{equation}{section}
\numberwithin{theorem}{section}

\def\qed{\hfill\hbox{${\vcenter{ \vbox{\hrule height 0.4pt\hbox{\vrule width
0.4pt height 6pt \kern5pt\vrule width 0.4pt}\hrule height 0.4pt}}}$}}
\newcommand{\mc}[1]{{\mathcal #1}}
\newcommand{\bb}[1]{{\mathbb #1}}

\begin{document}

\title{Directed current in quasi-adiabatically ac-driven nonlinear 
systems}

\author{Paolo Butt\`a\footnote{Dipartimento di Matematica, SAPIENZA Universit\`a di Roma, P.le Aldo Moro 2, 00185 Roma, Italy. E-mail: {\tt butta@mat.uniroma1.it, piero.negrini@uniroma1.it.}} \and Piero Negrini$^*$}

\maketitle

\begin{abstract} 
We rigorously prove the existence of directed transport for a certain class of ac-driven nonlinear one dimensional systems, namely the generation of transport with a preferred direction in the absence of a net driving force. 
\end{abstract}


   \section{Introduction}
   \label{sec:1}
   
In this paper we study the directed transport in ac-driven nonlinear one dimensional systems, whose dynamics is governed by an evolution equation of the following type:
\begin{equation}
\label{i0}
\ddot x + \gamma\dot x + U'(x) = E(t).
\end{equation}
This equation describes, for example, the classical motion of a particle in the potential $U(x)$ under the influence of friction ($\gamma>0$) and space-homogeneous external field $E(t)$. We assume $U(x)=U(x+2\pi)$ a space-periodic function and $E(t) = E(t+\mc T)$ a time-periodic function with zero mean. 

The paradigmatic case of \eqref{i0} is the non-conservative nonlinear pendulum, $U(x)=-\cos x$, which has been extensively studied. We recall in particular the comprehensive review \cite{Ma} by Mawhin on the forced (damped or undamped) pendulum, where one can find a large number of results on the existence of periodic solutions of first and second kind, on the existence of Mather sets, and on the chaotic behavior of solutions.
We also recall the paper \cite{Ca} by Casdagli, dedicated to the study of attractors of dissipative twist maps, where the stroboscopic map of the (periodically) forced damped nonlinear pendulum is given as the main example. 

The transport properties of \eqref{i0} with $\gamma\ge 0$ have been considered by several authors in the physical literature \cite{DYR,FYZ,JKH,M}. These numerical and theoretical studies show that if suitable space-time symmetries are broken then the system can sustain directed current. That is, an ensemble of trajectories with zero current at time $t=0$ produces a nonzero finite current as $t\to\infty$. 

The first paper  on this subject \cite{JKH} was motivated by previous work on stochastic ratchets, where directed transport of Brownian particles in asymmetric periodic potentials (ratchets) is induced by the application of non thermal forces. We refer to \cite{R} for an exhaustive review on Brownian motors. In \cite{JKH} the authors study the effect of finite inertia for the ratcheting mechanism in the absence of stochastic forces. By using \eqref{i0} with $E(t)$ harmonic and $U(x)$ a two-harmonic non symmetric potential as a working model, they investigate the role of regular and chaotic motion in directed transport; in particular, to what extent deterministically induced chaos resembles the role of noise. They showed numerically that the system can exhibit current flows, whose direction can be controlled by varying the amplitude of the external force. See also \cite{M}, where the origin of this current reversal is identified as a biforcation from a chaotic to a periodic regime.

The role of space-time symmetries is systematically analyzed in \cite{FYZ}. A necessary condition for the generation of transport is that some relevant space-time symmetries in \eqref{i0} are broken. In fact, nonzero average currents cannot take place in the presence of symmetries which allow to generate out of a given trajectory another one with reversed velocity. In the dissipative case $\gamma>0$ this happens if the potential $U(x)$ is an even function and the driving force $E(t)$ is antisymmetric around $\mc T/2$. In fact, with these hypothesis, \eqref{i0} is invariant under the symmetry $(x,t) \to (-x,t+\mc T/2)$. On the contrary, directed current may occur if the above symmetry is violated. This can be realized by means of an asymmetric driving force (\textit{underdamped asymmetrically tilting ratchet}) or by considering an asymmetric potential (\textit{underdamped rocking ratchet}). In fact, directed current is numerically observed in \cite{FYZ}, where, in the case of small dissipation, the rectification mechanism is explained in terms of a desymmetrization of the basins of attraction of two limit cycles with opposite velocities.

Purpose of the present paper is to rigorously prove the generation of directed current due to broken space-time symmetry in \eqref{i0}. Due to the complexity of the dynamics, it is not conceivable to obtain results in the most general case. We instead restrict our analysis to the simplest situation when the changes of the external field $E(t)$ in the course of time are extremely slow with respect to the characteristic time scales of the internal (non-driven) dynamics. In this regime, the system is expected to be well described by the so-called \textit{adiabatic approximation}, in which the current at time $t$ has essentially the same value of the steady state current $J_E$ corresponding to a static external field $E=E(t)$, the time $t$ merely playing the role of a parameter. In this approximation, the asymptotic time averaged current due to the periodic driving $E(t)$ is simply given by $\frac 1{\mc T}\int_0^{\mc T}\!dt\, J_{E(t)}$. 

But the above approximation is valid for single valued response functions (i.e.\ not containing hysteretic loops). In the case of \eqref{i0} this requirement turns out to be satisfied if $|E|>\max|U(x)|$, since for such static fields the corresponding autonomous system exhibits a unique globally attractive limit cycle. Then, in the case of  slowly varying, time-periodic external field which fulfills the above condition for large part of the time (see the next section for the precise definition), we obtain upper and lower bound on the time averaged current by means of its adiabatic approximation. However, this approximation is still too complicated to infer the sign of the current, even in special simple cases. We then restrict ourselves to the case of strongly enough driving force, which allows us to predict the direction of the current in typical examples of both asymmetrically tilting and rocking ratchets. 

The plan of the paper is the following: the notation and results are stated in Section~\ref{sec:2}, the steady state current is computed in Section~\ref{sec:3}, and the adiabatic limit is proved in Section~\ref{sec:4}. Finally, Section~\ref{sec:5} is devoted to the asymptotic expansion of the current for strong external force.

   \section{Notation and statement of the result}
   \label{sec:2}
  
Hereafter in the paper we consider the equation
\begin{equation}
\label{p1}
\ddot x + \gamma\dot x + U'(x) = E_\delta(\lambda t). 
\end{equation} 
The potential $U(x)$ is a regular non-constant periodic function of period $2\pi$ and we set
\begin{equation}
\label{p2}
- m := \min_{x\in [0,2\pi]}U'(x), \quad 
M := \max_{x\in [0,2\pi]}U'(x)
\end{equation}
(clearly $m,M >0$). The positive parameter $\lambda$ appearing in \eqref{p1} gives the ratio between the time scale of the autonomous non-driven system and the one of the driving force 
$E_\delta(\tau)$, which is chosen to be a mollified version of a time-periodic, mean zero, piecewise constant function. More precisely, $E_\delta(\tau)$ is the periodic function of period $T$ such that
\begin{equation}
\label{p3}
E_\delta (\tau) = \begin{cases}
{\displaystyle \frac 12 (E_1-E_2) + \frac 12 (E_1+E_2)
\phi_\delta(\tau-T_\delta)} & 
\text{if $\tau\in [0,T_\delta+\delta]$}, \\ \\
{\displaystyle \frac 12 (E_1-E_2) - \frac 12 (E_1+E_2) 
\phi_\delta(\tau-T+\delta)} & \text{if $\tau\in (T_\delta+\delta,
T]$}, \end{cases}
\end{equation}
where $0<2\delta<T-T_\delta$, $\phi_\delta(\tau) = \phi(\tau/\delta)$ with $\phi(s)$ a smooth decreasing function such that $\phi(s)=1$, respectively, $\phi(s) = -1$, if $s\le 0$, respectively, $s\ge 1$. The time $T_\delta$ is chosen so that $E_\delta(\cdot)$ has zero average,
\begin{equation*}
\frac{1}{T}\int_0^{T}\!d\tau E_\delta(\tau) = 0.
\end{equation*}
Note that this implies
\begin{equation}
\label{p3b}
T_0 = \lim_{\delta\to 0}T_\delta = \frac{E_2}{E_1+E_2}\; T.
\end{equation}

In order to state precisely our result, first we briefly discuss the qualitative behavior of the autonomous equation (static $E$),
\begin{equation}
\label{p5}
\ddot x + \gamma\dot x + U'(x) = E.
\end{equation} 
We denote by $v$ the velocity of the system, the pair $(x,v)$ thus giving coordinates in the cylindrical phase space $S^1\times\bb R$. More precisely, we consider the development of the cylinder on the $(x,v)$ plane. Equation \eqref{p5} is then equivalent to the first order system 
\begin{equation}
\label{p6}
\left\{\begin{array}{l} \dot x = v, \\ \dot v = E - U'(x) - \gamma v. \end{array}\right.
\end{equation} 
Note that the equation of the corresponding integral curves on the cylinder is
\begin{equation}
\label{p7}
\frac{dv}{dx} = - \gamma + \frac{E-U'(x)}{v}.
\end{equation}

\begin{proposition}
\label{cili}
Recalling \eqref{p2}, the following holds. If $E>M$, respectively, $E<-m$, there exists a unique globally attractive limit cycle encircling the cylinder, which lies in the $v>0$, respectively, $v<0$ region. We denote the cycle by $v_E(x)$, that is $v_E(x)$ turns out to be the unique periodic solution to \eqref{p7}, and let $\mc V_E$ be the corresponding mean velocity, that is
\begin{equation}
\label{p8}
\mc V_E = 2\pi \left[\int_0^{2\pi}\!dx\, \frac{1}{v_E(x)}\right]^{-1}. 
\end{equation}
Then, there exists $\bar C >0$ such that, if $x(t;x_0,v_0)$ is the solution to \eqref{p5} with initial conditions $(x_0,v_0) \in S^1\times\bb R$,
\begin{equation}
\label{p9}
\left|\frac{x(t;x_0,v_0)}t -\mc V_E\right| \le \bar C \;\frac{(1+|v_0|)\log(e+|v_0|)}t.
\end{equation}
\end{proposition}

An appropriate tool for describing transport is to analyze the statistical solutions of the dynamics, that is the collective behavior of an ensemble of independent particles, whose evolution is governed by the equation of motion \eqref{p1}. We denote by $X_{\lambda,\delta}(t;x_0,v_0)$ the solution to \eqref{p1} with initial conditions $X_{\lambda,\delta}(0;x_0,v_0)=x_0$, $\dot X_{\lambda,\delta}(0;x_0,v_0) = v_0$. Given any initial probability measure $\mu(dx_0,dv_0)$ on the cylindrical phase space $S^1\times\bb R$, the {\it current} at time $t$ is defined by averaging the drift of a single trajectory over the initial conditions:
\begin{equation}
\label{pp1}
J_{\lambda,\delta}(t) = \int_{S^1\times\bb R}\!\mu(dx_0\, dv_0)\; \frac{X_{\lambda,\delta}(t;x_0,v_0)}t.
\end{equation}
We say that the system exhibits directed transport if $J_{\lambda,\delta}(t)$ is definitively away from zero as $t\to\infty$. We will show that this is the case for suitable choices of the potential and of the driving force, provided the parameters $\lambda$ and $\delta$ are sufficiently small. More precisely, we first prove that, under further assumptions on the driving force, the {\it adiabatic limit} $\lambda\to 0$ of the current can be explicitly computed. We then show the appearance of nonzero currents in this limit for suitable choices of the potential and driving force. By continuity, small deviation from the adiabatic limit will change numbers but not the fact of nonzero currents.

\begin{theorem}
\label{adli}
Recall the definition \eqref{pp1} and assume that
\begin{equation}
\label{p4}
E_1>M,\quad E_2 > m, 
\end{equation}
\begin{equation}
\label{p4b}
\int_{S^1\times\bb R}\!\mu(dx_0\, dv_0) \; (1+|v_0|)\log(e+|v_0|) < \infty.
\end{equation}
Then there exists the limit
\begin{equation}
\label{adliebis}
J = \lim_{\delta\to 0} \liminf_{\lambda\to 0}\liminf_{t\to\infty} J_{\lambda,\delta}(t) \; = \; \lim_{\delta\to 0} \limsup_{\lambda\to 0}\limsup_{t\to\infty} J_{\lambda,\delta}(t). 
\end{equation}
Moreover
\begin{equation}
\label{adlie}
J = \frac{E_2\mc V_{E_1} + E_1\mc V_{-E_2}}{E_1+E_2}.
\end{equation} 
\end{theorem}

   \section{The autonomous case}
   \label{sec:3}
   
In this section we prove Proposition~\ref{cili}, see also \cite[Chapter VII, \S 3]{AVK}, where the particular case of the non-conservative pendulum with constant torque is treated. We prove the claim for $E>M$. The case $E<-m$ can be deduced by the previous one by noticing that $y(t) := -x(t;x_0,v_0)$ solves the equation
\begin{equation*}
\ddot y + \gamma\dot y  + \tilde{U}'(y) = - E
\end{equation*}
with initial conditions $(-x_0,-v_0)$ and potential $\tilde U(y) = U(-y)$. 

First of all we observe that the divergence of the vector field on the right hand side of \eqref{p6} is equal to $-\gamma<0$. Therefore, by Green's theorem, the system \eqref{p6} does not have any closed paths not encircling the cylinder and can have at most one limit cycle encircling the cylinder. We now prove that such a limit cycle does exist, it is globally attractive, and lies entirely on the upper half $v>0$ of the cylinder. 

Given $(x_0,v_0)\in S^1\times\bb R$, we use in the sequel the abbreviate notation
\begin{equation}
\label{p10}
x(t) = x(t;x_0,v_0), \quad v(t) = \dot x(t;x_0,v_0).
\end{equation}
By Duhamel's formula,
\begin{equation*}
v(t) = e^{-\gamma t} v_0 + \int_0^t\!ds\, e^{-\gamma(t-s)} \big[E-U'(x(s))\big],
\end{equation*}
whence
\begin{equation*}
\frac{E-M}\gamma + \left(v_0-\frac{E-M}\gamma\right) 
e^{-\gamma t} \le v(t) \le \frac{E+m}\gamma + \left(v_0-\frac{E+m}\gamma\right) e^{-\gamma t}. 
\end{equation*}
It follows that for any $\alpha\ge 0$ the strip 
\begin{equation*}
\mc R_\alpha := S^1\times \mc I_\alpha, \qquad \mc I_\alpha := \left[\frac{E-M}\gamma-\alpha,\frac{E+m}\gamma + \alpha \right]
\end{equation*}
is positively invariant. In particular, for a suitable number $C_0>0$,
\begin{equation}
\label{p10b}
|v(t)| \le  C_0 (1+|v_0|) \qquad \forall\, t \ge 0.
\end{equation}
Moreover, for each $\alpha>0$ there exists a positive number $C_1=C_1(\alpha)$ such that 
\begin{equation}
\label{p11}
((x(t),v(t)) \in \mc R_\alpha \qquad \forall\, t \ge C_1\log(e+|v_0|).
\end{equation}
Since $E>M$ there are no singular points, therefore the Poincar\'e-Bendixson theorem implies the existence of a limit cycle inside the strip $\mc R_\alpha$. We have already shown that this cycle is unique, and therefore asymptotically stable with $\mc R_\alpha$ inside its basin of attraction. By \eqref{p11} we conclude that the cycle is globally attractive. We denote by $v_E(x)$ the corresponding integral curve, which is therefore the unique periodic solution to \eqref{p7}. 

It remains to prove the estimate \eqref{p9}. We fix $\alpha<\gamma^{-1}(E-M)$ and consider the Poincar\'e map 
\begin{equation*}
G : \mc I_\alpha \to \mc I_\alpha
\quad : \quad G(v) := \Phi(2\pi;v),
\end{equation*}
where $\Phi(x;v)$ denotes the solution to \eqref{p7} with initial condition $\Phi(0;v)=v$. Clearly $v^*:=v_E(0)$ is the unique fixed point of $G$. Since 
\begin{equation*}
\frac{d}{dx}\frac{\partial \Phi(x;v)}{\partial v} = - \frac{E-U'(x)}{\Phi(x;v)^2}\;\frac{\partial \Phi(x;v)}{\partial v},\qquad \frac{\partial \Phi(0;v)}{\partial v} = 1,
\end{equation*}
then
\begin{equation*}
\frac{\partial \Phi(x;v)}{\partial v} = \exp\left[-\int_0^x\!dy\, \frac{E-U'(y)}{\Phi(y;v)^2}\right] = \frac v{\Phi(x;v)}\exp\left[- \int_0^x\!dy\, \frac{\gamma}{\Phi(y;v)}\right],
\end{equation*}
where in the last equality we used again that $\Phi(x;v)$ solves \eqref{p7}. Therefore
\begin{equation*}
G'(v^*) = \frac{\partial \Phi(2\pi;v)}{\partial v}\bigg|_{v=v^*} = 
\exp\left[- \int_0^{2\pi}\!dx\, \frac{\gamma}{v_E(x)}\right].
\end{equation*}
Since $v_E(\cdot) >0$ then $0<G'(v^*)<1$, whence there exist $C_2, \beta >0$ such that
\begin{equation}
\label{p12}
\big|G^k(v) - v^*\big| \le C_2 e^{- \beta k} \qquad \forall\, v\in \mc I_\alpha \quad\forall\, k\in \bb N.
\end{equation}
Recalling the notation \eqref{p10}, we denote by $t_0,t_1,t_2,\ldots$, the increasing sequence of times such that $x(t_k) = 0 \textrm{ (mod }2\pi)$ and $v(t_k)\in \mc I_\alpha$ (clearly $x(t_k)-x(t_{k-1})=2\pi$). Given $t>0$ let $n=n(t)= \max\{k\ge 0 : t_k<t\}$. By \eqref{p10b} and \eqref{p11}, since $\mc R_\alpha$ is contained in the upper half $v>0$ of the cylinder and $x_0\in [0,2\pi)$, there exists $C_3>0$ such that 
\begin{equation}
\label{p13}
|x(t_0)| \le C_3 (1+|v_0|) \log(e+|v_0|), \qquad t_0 \le C_3\log(e+|v_0|)
\end{equation}
and
\begin{equation}
\label{p14}
t-t_n \le C_3, \qquad x(t)-x(t_n) \le 2\pi. 
\end{equation}
Noticing that $x(t_n)-x(t_0) = 2\pi n$ we have
\begin{eqnarray}
\label{p15}
\left|\frac{x(t)}t - \mc V_E \right| & \le & \frac{|x(t)-x(t_n)| + |x(t_0)| + \mc V_E (t-t_n+t_0)} t \nonumber \\ && + \frac{\mc V_E}t\left| \frac{2\pi n}{\mc V_E} - (t_n-t_0)\right|,
\end{eqnarray}
where
\begin{eqnarray*}
\frac{2\pi n}{\mc V_E} - (t_n-t_0) & = &
\sum_{k=1}^n \left[\frac{2\pi}{\mc V_E} - (t_k-t_{k-1})\right]
\\ & = & \sum_{k=1}^n \int_0^{2\pi}\!dx \left[\frac{1}{v_E(x)} - \frac{1}{\Phi(x;v(t_{k-1}))} \right].
\end{eqnarray*}
Since $v(t_{k-1}) = G^{k-1}(v(t_0))$, by \eqref{p12} and the uniform continuity of $\Phi(x;v)$ in $\mc R_\alpha$, we get, for some $C_4>0$,
\begin{equation}
\label{p16}
\left| \frac{2\pi n}{\mc V_E} - (t_n-t_0)\right| \le C_4.
\end{equation}
Plugging the estimates \eqref{p13}, \eqref{p14}, and \eqref{p16} into 
\eqref{p15} we get \eqref{p9} for a suitable positive constant $\bar C$. 

  \section{The adiabatic limit}
  \label{sec:4}
   
In this section we prove Theorem~\ref{adli}. By the same reasoning of the previous section, using Duhamel's formula, we conclude that there exists a positive number $A$ (independent of $\lambda$ and $\delta$) such that
\begin{equation}
\label{p17}
\big|\dot X_{\lambda,\delta}(t;x_0,v_0)\big|  \le A (1+|v_0|)
\qquad\forall\, t\ge 0.
\end{equation}
Now we define, for any integer $k\ge 0$,
\begin{equation*}
t_k^{(1)} = \lambda^{-1} kT, \qquad
t_k^{(2)} = \lambda^{-1} (kT + T_\delta),
\end{equation*}
\begin{equation*}
t_k^{(3)} = \lambda^{-1} (kT + T_\delta+\delta),\qquad
t_k^{(4)} = \lambda^{-1} (kT + T-\delta),
\end{equation*}
and denote by $N=N(t)$ the largest integer $k$ for which $t_k^{(1)} < t$. In the sequel we shorthand $X_{\lambda,\delta}(t;x_0,v_0)$ by $X(t)$. We decompose
\begin{equation*}
X(t) =  X(t) - X(t_N^{(1)}) + \sum_{k=0}^{N-1} \sum_{i=1}^4 \left[ X(t_k^{(i+1)}) - X(t_k^{(i)})\right],
\end{equation*}
with the convention $t_k^{(5)}=t_{k+1}^{(1)}$. Since $t-t_N^{(1)}\le\lambda^{-1}T$ and $t_k^{(3)} - t_k^{(2)} = t_k^{(5)}-t_k^{(4)} = \lambda^{-1}\delta$, by \eqref{p17} we have that
\begin{equation*}
\big|X(t) - X(t_N^{(1)})\big| \le A(1+|v_0|)\lambda^{-1}T, 
\end{equation*}
\begin{equation*}
\big|X(t_k^{(3)}) - X(t_k^{(2)})\big| + \big|X(t_k^{(5}) - X(t_k^{(4)})\big| \le 2A(1+|v_0|) \lambda^{-1}\delta.
\end{equation*}
On the other hand, $E_\delta(\lambda t) = E_1$ for $t \in [t_k^{(1)},t_k^{(2)}]$ and $E_\delta(\lambda t) = - E_2$ for $t \in [t_k^{(3)},t_k^{(4)}]$. Therefore, since $t_k^{(2)} - t_k^{(1)}= \lambda^{-1}T_\delta$, $t_k^{(4)} - t_k^{(3)} = \lambda^{-1}(T-T_\delta-2\delta)$, and recalling the hypothesis \eqref{p4}, by \eqref{p9} and \eqref{p17} it follows that
\begin{equation*}
\big|X(t_k^{(2)}) - X(t_k^{(1)}) - \lambda^{-1}T_\delta\mc V_{E_1}\big|
\le \bar C_1 \, (1+A+A|v_0|)\log(e+A+A|v_0|), 
\end{equation*}
\begin{equation*}
\big|X(t_k^{(4)}) - X(t_k^{(3)}) - \lambda^{-1}(T-T_\delta-2\delta)\mc V_{-E_2}\big| \le \bar C_2 \, (1+A+A|v_0|)\log(e+A+A|v_0|),
\end{equation*}
where $\bar C_1$, respectively, $\bar C_2$ are equal to the number $\bar C=\bar C(E)$ appearing in \eqref{p9} for $E=E_1$, respectively, $E=-E_2$. Collecting all the previous estimates we conclude that, for some number  $C>0$,
\begin{eqnarray*}
&& \left|\frac{X(t)}t - \frac{N}{\lambda t}\left[T_\delta\mc V_{E_1}+(T-T_\delta-2\delta)\mc V_{-E_2}\right]\right| \\ && ~~~~~~~~~~~~~~ \le C \frac{N}{\lambda t}\left(\lambda+\delta+\frac 1N\right) (1+|v_0|)\log(e+|v_0|).
\end{eqnarray*}
Noticing $N(\lambda t)^{-1}\to T^{-1}$ as $t\to+\infty$ and recalling \eqref{p3b}, by assumption \eqref{p4b} the claims \eqref{adliebis} and \eqref{adlie} follow.

  \section{Currents}
  \label{sec:5}

In this section we show how directed current actually occurs for suitable choices of the driving force and/or the potential. As discussed in the Introduction, a necessary condition for the generation of transport is that the space-time symmetry $(x,t) \to (-x,t+\lambda^{-1}T/2)$ in equation \eqref{p1} is broken. This symmetry is satisfied if the potential $U(x)$ is an even function and the driving force $E_\delta(\tau)$ is antisymmetric around $T/2$ (this is obtained, see \eqref{p3}, by choosing $E_1=E_2$, $T_\delta=(T-\delta)/2$, and $\phi(s)$ antisymmetric around $1/2$). It can be violated by means of an asymmetric driving force or by considering an asymmetric potential. Since the adiabatic approximation \eqref{adlie} is not explicitly known for generic potentials, we cannot establish the direction of the current for any value of the parameters. We instead expand this expression  for large values of the driving force, in order to answer the question in at least some special cases. More precisely, we set $E_1=E+\Delta$, $E_2=E$ and perform an asymptotic expansion of the current for diverging values of $E$.

Let us first rewrite the expression of $J$ in a more useful form. Recalling $\Phi(x;v)$ denotes the solution to \eqref{p7} with initial condition $\Phi(0;v)=v$, we note that $\tilde\Phi(x,v) := -\Phi(-x;v)$ solves
\begin{equation*}
\frac{dv}{dx} = - \gamma + \frac{-E-\tilde U'(x)}{v}, \qquad
\tilde U(x) := U(-x),
\end{equation*}
with initial condition $\tilde\Phi(0,v) = -v$. In particular, for $E>m$ the unique periodic solution to the above equation is $\tilde v_E(x) := - v_{-E}(-x)$. Therefore, denoting by $\tilde{ \mc V}_E$ the mean velocity on the cycle $\tilde v_E(x)$, the current reads
\begin{eqnarray*}
J & = & \frac{E\mc V_{E+\Delta} - (E+\Delta)\tilde{ \mc V}_{E}}{2E+\Delta} \\ & = & \frac{\mc V_{E+\Delta} \tilde{\mc V}_{E}}{2\pi(2E+\Delta)}\left[ \int_0^{2\pi}\! dx \frac{E}{\tilde v_{E}(x)} - \int_0^{2\pi}\! dx \frac{E+\Delta}{v_{E+\Delta}(x)}\right].
\end{eqnarray*}

Now we need an asymptotic expansion of the limit cycle for $E$ large. Recall $v_E(x)$ is a solution to \eqref{p7}, so that
\begin{equation}
\label{p18}
\frac 12 \frac{d}{dx} v_E(x)^2 = -\gamma v_E(x) + E - U'(x). 
\end{equation}
Integrating along the circle, since $v_E$ and $U'$ are periodic, we obtain
\begin{equation*}
\frac{1}{2\pi} \int_0^{2\pi}\!dx\, v_E(x) = \frac E\gamma.
\end{equation*}
For any fixed integer $N\ge 1$ and $E$ large enough, we then look for an expansion of the following type,
\begin{equation}
\label{exp}
v_E(x) = \frac E\gamma + \sum_{k=1}^N \frac{v_k(x)}{E^k} + \frac{R_{N,E}(x)}{E^{N+1}},
\end{equation}
where $v_k(x)$ and $R_{N,E}(x)$ are periodic functions with zero average and $R_{N,E}(x)$ is uniformly bounded as $E$ diverges. By inserting this expansion in \eqref{p18} and equating to zero the first $N-1$ powers of $1/E$, we get that the set of functions $\{v_k(x)\}$ must be solution to the following linear differential system in triangular form, 
\begin{equation}
\label{p19}
\left\{\begin{array}{l} 
v_1'(x) = -\gamma U'(x), \\ \\
v_2'(x) = -\gamma^2 v_1(x), \\
v_{k+1}'(x) = - \gamma^2 v_k(x) - {\displaystyle \gamma\sum_{\ell=1}^{k-1}v_\ell'(x)v_{k-\ell}(x) \qquad k=3,\ldots, N-1.} \end{array}\right.
\end{equation}
Once the functions $v_k(x)$ are computed, the remainder $R_{N,E}(x)$ is  determined by imposing that the right hand side of \eqref{exp} is solution to \eqref{p18}. By a standard application of the Gronwall lemma, it is easy to verify that $R_{N,E}(x)$ turns out to be uniformly bounded as $E$ diverges, we omit the details. Analogously to $v_E(x)$, the function $\tilde v_E(x)$ admits the expansion
\begin{equation*}
\tilde v_E(x) = \frac E\gamma + \sum_{k=1}^N \frac{\tilde v_k(x)}{E^k} + \frac{\tilde R_{N,E}(x)}{E^{N+1}},
\end{equation*}
where the set of functions $\{\tilde v_k(x)\}$ satisfy the same system \eqref{p19} with $\tilde U'(x)$ instead of $U'(x)$ in the first equation. We note that this implies
\begin{equation}
\label{p199}
\tilde v_k(x) = \left\{\begin{array}{ll} v_k(-x) & \text{ if $k$ is odd,} \\ -v_k(-x) & \text{ if $k$ is even.} \end{array} \right.
\end{equation}

In the sequel, to simplify the notation, we shall assume, without loss of generality, that $U(x)$ has zero average. By using the above expansions for $N=1$ and recalling $v_1(x)$ has zero mean, we have
\begin{eqnarray*}
\int_0^{2\pi}\! dx \frac{1}{v_{E}(x)} & = & \frac \gamma E\int_0^{2\pi}\!dx\, \frac 1{1+\frac\gamma {E^2} \left(v_1(x) + \frac 1E R_{1,E}(x) \right)} \\ & = & \frac{2\pi\gamma}E + \frac{\gamma^3}{E^5}\int_0^{2\pi}\!dx\,v_1(x)^2 + O\left(\frac 1{E^7}\right).
\end{eqnarray*}
Analogously,
\begin{eqnarray*}
\int_0^{2\pi}\! dx \frac{1}{\tilde v_{E}(x)} = \frac{2\pi\gamma}E + \frac{\gamma^3}{E^5}\int_0^{2\pi}\!dx\,\tilde v_1(x)^2 + O\left(\frac 1{E^7}\right).
\end{eqnarray*}
Now, by \eqref{p19} and \eqref{p199}, $v_1(x) = - \gamma U(x)$ and
$\tilde v_1(x) = v_1(-x)$, whence
\begin{equation*}
\int_0^{2\pi}\!dx\,\tilde v_1(x)^2 = \int_0^{2\pi}\!dx\, v_1(x)^2 = \gamma^2 \int_0^{2\pi}\!dx\,U(x)^2,
\end{equation*}
which is positive provided $U(x)$ is not identically zero. Then   
\begin{equation*}
J = \frac{\mc V_{E+\Delta} \tilde{\mc V}_{E}\gamma^5}{2\pi(2E+\Delta)}\left[\left(\frac{1}{E^4} - \frac{1}{(E+\Delta)^4}\right)\int_0^{2\pi}\!dx\,U(x)^2 + O\left(\frac 1{E^7}\right)\right],
\end{equation*}
which implies that for $E$ sufficiently large the current has the sign of $\Delta$.

If instead $\Delta=0$ the current reads
\begin{equation*}
J = \frac{\mc V_E \tilde{\mc V}_E}{4\pi} \int_0^{2\pi}\!dx  \left[\frac{1}{\tilde v_E(x)} - \frac{1}{v_E(x)}\right].
\end{equation*}
By the expansion of $v_E(x)$ and $\tilde v_E(x)$ for $N=4$ and using that $v_k(x),\tilde v_k(x)$ have zero mean, again by \eqref{p199}, we have
\begin{eqnarray*}
&& \int_0^{2\pi}\!dx  \left[\frac{1}{\tilde v_E(x)} - \frac{1}{v_E(x)}\right] =\int_0^{2\pi}\!dx  \left[\frac{1}{\tilde v_E(-x)} - \frac{1}{v_E(x)}\right] \\ && ~~~~~~~~~~~~~~~~~~~~ = \, - \frac{4\gamma^3}{E^6}\int_0^{2\pi}\!dx\,  v_1(x)\,v_2(x) + \frac{6\gamma^4}{E^8}\int_0^{2\pi}\!dx\,  v_1(x)^2\,v_2(x)  \\ && ~~~~~~~~~~~~~~~~~~~~~~~~ - \, \frac{4\gamma^3}{E^8}\int_0^{2\pi}\!dx\, \big[v_2(x)\,v_3(x) + v_1(x)\,v_4(x)\big] + O\left(\frac 1{E^{10}}\right).
\end{eqnarray*}
Now, by \eqref{p19}, 
\begin{equation*}
\int_0^{2\pi}\!dx\,  v_1(x)\,v_2(x) = -\frac{1}{\gamma^2} \int_0^{2\pi}\!dx\,  v_2'(x)\,v_2(x) = 0,
\end{equation*}
\begin{equation*}
\int_0^{2\pi}\!dx\,  v_1(x)^2\,v_2(x) = \frac{1}{\gamma^4}\int_0^{2\pi}\!dx\, v_2'(x)^2\, v_2(x),
\end{equation*}
and, integrating by parts,
\begin{eqnarray*}
\int_0^{2\pi}\!dx\, v_1(x)\,v_4(x) & = & - \frac{1}{\gamma^2}
\int_0^{2\pi}\!dx\, v_2'(x)\,v_4(x) = \frac{1}{\gamma^2}
\int_0^{2\pi}\!dx\, v_2(x)\,v_4'(x) \\ & = & 
- \frac{1}{\gamma^2} \int_0^{2\pi}\!dx\,  v_2(x)\,\big[\gamma^2 v_3(x)+ \gamma\, (v_1v_2)'(x)\big] \\ & = & 
- \int_0^{2\pi}\!dx\, v_2(x)\,v_3(x) + \frac 1\gamma
\int_0^{2\pi}\!dx\, v_1(x)\,v_2(x)\,v_2'(x)
\\ & = &  - \int_0^{2\pi}\!dx\, v_2(x)\,v_3(x) - \frac 1{\gamma^3}
\int_0^{2\pi}\!dx\, v_2(x)\,v_2'(x)^2.
\end{eqnarray*}
Again by \eqref{p19}, $v_2(x) = \gamma^3 G(x)$ where $G(x)$ is the antiderivative of $U(x)$ with zero average, whence 
\begin{equation*}
J = \frac{5\mc V_E \tilde{\mc V}_E \gamma^9}{2\pi E^8}\left[ \int_0^{2\pi}\!dx\, G(x)\,G'(x)^2 + O\left(\frac 1{E^2}\right)\right].
\end{equation*}
The integral on the right hand side is nonzero for a generic not even potential. Let us determine the direction of the current in two typical examples of ratchet potential \cite{R}.

\textit{i)} Assume $U(x)$ is a piecewise linear ``sawtooth'' type potential, that is
\begin{equation*}
U(x) = \left\{\begin{array}{ll} {\displaystyle \frac{bx}a - \frac b2} & \text{ if } x\in [0,a], \\ \\ {\displaystyle \frac{b(2\pi-x)}{2\pi-a} - \frac b2} & \text{ if }x\in (a,2\pi], \end{array} \right.
\end{equation*}
where $a\in (0,2\pi)$, $b>0$, and the constant term $-b/2$ makes $U$ of zero average (clearly, the equation of motion has to be thought with some mollified version of $U$). The antiderivative of $U(x)$ with zero average is 
\begin{equation*}
G(x) = \left\{\begin{array}{ll} {\displaystyle \frac{bx^2}{2a}  - \frac{bx}2 - \frac{b\pi(\pi-a)}3} & \text{ if } x\in [0,a], \\ \\ {\displaystyle - \frac{b(2\pi-x)^2}{2(2\pi-a)} + \frac{b(2\pi-x)}2  - \frac{b\pi(\pi-a)}3} & \text{ if }x\in (a,2\pi].\end{array} \right.
\end{equation*}
Then, integrating by parts, 
\begin{equation*}
\int_0^{2\pi}\!dx\, G(x)\,G'(x)^2 = - \frac 12  
\int_0^{2\pi}\!dx\, G(x)^2\,G''(x) = - I_1 + I_2,
\end{equation*}
where
\begin{eqnarray*}
I_1 & = & \frac{b^3}{2a} \int_0^a\!dx\, \left[\frac{x^2}{2a} - \frac x2 - \frac{\pi(\pi-a)}3 \right]^2, \\
I_2 & = & \frac{b^3}{2(2\pi-a)} \int_a^{2\pi}\!dx\, \left[- \frac{(2\pi-x)^2}{2(2\pi-a)} +\frac{2\pi-x}2 - \frac{\pi(\pi-a)}3 \right]^2.
\end{eqnarray*}
With the substitution $x = a + y(2\pi-a)/a$ the integral $I_2$ becomes
\begin{equation*}
I_2 = \frac{b^3}{2a} \int_0^a\!dy\, \left[\frac{2\pi-a}a\left(\frac{y^2}{2a} - \frac y2\right) + \frac{\pi(\pi-a)}3 \right]^2,
\end{equation*}
so that
\begin{equation*}
\int_0^{2\pi}\!dx\, G(x)\,G'(x)^2 = \frac{b^3\pi(\pi-a)}{a^2} \int_0^a\!dx\,\left(\frac{x^2}{2a}-\frac x2\right) \left(\frac{x^2}{a^2}-\frac xa +\frac{2\pi}3\right).
\end{equation*}
Since the integrand is negative for any $x\in (0,a)$, we conclude that the current is positive for $a>\pi$ and negative for $a<\pi$. In other words, the current has the sign of the flatter slope of $U(x)$. 

\textit{ii)} Assume $U(x)$ is a two-harmonic potential, that is 
\begin{equation*}
U(x) = \sin x + \mu \sin(2x), \qquad 0<\mu<1,
\end{equation*}
whence
\begin{equation*}
G(x) = -\cos x - \frac \mu 2  \cos(2x).
\end{equation*}
By an explicit computation, 
\begin{equation*}
\int_0^{2\pi}\!dx\, G(x)\,G'(x)^2 = - \frac{3\pi}4 \mu.
\end{equation*}
The current is negative, also in this case it goes into the direction of the steeper slope of $U(x)$. 

We remark that in the previous examples the current has the same direction of that appearing in the analogous cases of overdamped, adiabatically rocked ratchets in the zero temperature limit \cite{R}.

\section*{Acknowledgments}

The authors acknowledge the partial support of the italian MIUR (Cofin 2006). P.\ Butt\`a is grateful to the Institute Henri Poincar\'e - Centre Emile Borel in Paris for the hospitality, where part of this work has been done during his participation in the program ``Interacting particle systems, statistical mechanics and probability theory''.


\begin{thebibliography}{99}

\bibitem{AVK} A.A. Andronov, A.A. Vitt, and S.E. Khaikin. 
\textit{Theory of oscillators}.  
Dover Publications, Inc., New York, 1987.

\bibitem{Ca} M. Casdagli, Rational chaos  in dissipative systems, \textit{Phys. D} \textbf{3}, 365--386 (1988).

\bibitem{DYR} S. Flach, O. Yevtushenko, and K. Richter, Ac-driven phase-dependent directed diffusion, \textit{Phys. Rev. E} \textbf{61}, 7215 (2000).

\bibitem{FYZ} S. Flach, O. Yevtushenko, and Y. Zolotaryuk, Directed current due to broken time-space symmetry, \textit{Phys. Rev. Lett.} \textbf{84}, 2358--2361 (2000).

\bibitem{JKH} P. Jung, J.G. Kissner, and P. H\"anggi, Regular and chaotic transport in asymmetric periodic potentials: inertia ratchets,
\textit{Phys. Rev. Lett.} \textbf{76}, 3436--3439 (1996).

\bibitem{M} M.O. Mateos, Chaotic transport and current reversal in deterministic ratchets, \textit{Phys. Rev. Lett.} \textbf{84}, 258--261  (2000).

\bibitem{Ma} J. Mawhin, Global results for Forced Pendulum Equation, Handbook of Differential Equations, Ordinary Differential Equations \textbf{1}, 535--585. Edited by A. Ca\~{n}ada, P. Dr\'abek, and A. Fonda, Elsevier (2004).

\bibitem{R} P. Reimann, Brownian motors: noisy transport far from equilibrium, \textit{Phys. Rep.} \textbf{361}, 57--241 (2002).

\end{thebibliography}
\end{document}